\documentstyle[prd,aps]{revtex}

\begin{document}
\draft

 \title{\bf   The constant mean curvature slices of\\
asymptotically flat spherical  spacetimes}
\author{Mirta Iriondo$^*$, Edward Malec$^{**}$ and Niall \'O Murchadha$^{+}$}
\address{$^*$ Royal Institute of Technology, Mathematics Department, Stockholm,
Sweden}

\address{$^{**}$  Institute of Physics,  Jagellonian University,
30-059  Cracow, Reymonta 4, Poland}
\address{$^{+}$ Physics Department, University College, Cork,
Ireland}

\maketitle
\begin{abstract} We investigate the formation of trapped surfaces
in  asymptotically flat  spherical spacetimes, using constant mean curvature
slicing.
\end{abstract}
\pacs{04.20.Me, 95.30.Sf, 97.60.Lf, 98.80.Dr}

\section{INTRODUCTION}
\label{sec1}
In the analysis of General Relativity as a Hamiltonian system \cite{MTW} one
chooses a time function and considers the foliation of the spacetime by the
slices of constant time. Two natural geometrical quantities arise on such three
slices. One is the intrinsic three metric, usually $g_{ab}$, and the other is
the
extrinsic curvature $K^{ab}$, the time derivative of $g_{ab}$. These are not
independent: they are related by the constraints

\begin{eqnarray*}
{\cal R}^{(3)}- K^{ab}K_{ab}+(\text{tr} K)^2&=&16\pi\rho\\
\nabla_a K^{ab}- g^{ab}\nabla_a\text{tr} K=-8\pi j^a
\end{eqnarray*}
where ${\cal R}^{(3)}$ is the three scalar curvature, $\rho$ is the energy
density and $j^a$ is the current density of the sources.

It is often useful to specify the foliation, and thus the time, by placing a
condition on the extrinsic curvature, The most common choice in asymptotically
flat spacetimes is the maximal slicing condition, tr$K=0$. In cosmologies,
the favoured slicing is the constant mean curvature (CMC) foliation with
tr$K= constant$.

Such CMC slices have also been used in an asymptotically flat context
\cite{iriondo}.
They are everywhere spacelike, but at infinity they approach null infinity.
Thus they are very useful in investigating the relationship between spatial
and
null infinity. A standard model of CMC hypersurfaces are the mass hyperboloids
in Minkowski space \cite{chengyau}.

In this paper we investigate a very special class of CMC, those which are
spherically symmetric. Because of the absence of gravitational radiation,
spherical spacetimes are particularly simple, yet realistic, models of general
 solutions to the Einstein equations.

If we have a spherically symmetric three surface, the intrinsic metric can be
written as

$$
ds^2= dl^2+R^2 d\Omega^2
$$
where $l$ is the proper distance in the radial direction and $R$ is the
Schwarzschild
or areal radius. The geometry is  encoded into the relationship between $R$ and
$l$ and
a useful object to use is the mean curvature of the spherical two surfaces,
given by
$$
p={2 \over R} {dR\over dl}
$$
The constraints now can be written as
\begin{eqnarray}
\partial_l p&=& -8\pi \rho-{3\over 4}(K^r_r)^2- {3\over 4} p^2+{1\over R^2}
+{1\over 2}\text{tr}K K^r_r+{1\over 4}(\text{tr}K)^2\nonumber\\
\partial_l (K^r_r-\text{tr}K)&=& -{3\over 2}p K^r_r+{1\over 2}p\text{tr}K-
8\pi j_l
\label{constraint}
\end{eqnarray}

It has been recently shown \cite{EMNOM} that the constraints of General
Relativity in the spherically
symmetric case can be expressed very simply by using the null expansions as
subsidiary variables and the constraints can be expressed as a system of
quasilinear first order O.D.E.'s. We apply this new formulation in the CMC
case to investigate a number of interesting problems.

Much work has been carried out in recent years on how concentrations of matter
may gravitationally collapse \cite{EMNOM}, \cite{BEMNOMKOC}, \cite{zannias}.
One of the motivations for
repeating the calculation in various slicings of asymptotically flat
spacetimes is due to the fact that  no covariant formulation of the
question has been found. This article, in which  we derive both
necessary and sufficient conditions for the formation of trapped surfaces,
can be regarded as an attempt to see how the
criteria we obtain are more or less independent of the details of ths slicing
used. Let us emphasise that the appearance of trapped surfaces   indicates that
irreversible gravitational collapse  has commenced.

We   derive, for the sake of completeness,  the general line element for the
  Reissner-Nordstr\"om spacetime in the slicing by constant mean curvature
hypersurfaces.  That is a generalization of the corresponding solution
in the maximal slicing \cite{bryce}.

\section{CMC HYPERSURFACES IN MINKOWSKI SPACE. }
\label{sec2}

Let us consider  spherically symmetric CMC hypersurfaces in Minkowski
space. We write the four metric as

 \begin{equation}
ds^2 = -  d\tau^2 + \tau^2 [dr^2 +  \sinh^2 r  d\Omega^2],
\label{2.1}
\end{equation}
where $d\Omega^2 = d\theta^2 + \sin^2\theta d\phi^2$ is the standard
round two metric. The scalar curvature ${\cal R}^{(3)}$ of the three space
defined
by $\tau= constant$ is  ${\cal R}^{(3)}=-{6\over \tau^2}$.

 The   extrinsic curvature of this slice is pure trace, $K_{ab}\equiv {1\over
2}
 \partial_{\tau }g_{ab}= {1\over \tau}g_{ab} $ which  implies
tr$K=K={3\over \tau }$.
The proper radial distance $l$ along the slice is related to the coordinate
radius $r$ by
$\tau dr =dl$ which yields

\begin{eqnarray}
r={l\over \tau }={Kl\over 3}.
\label{2.2}
\end{eqnarray}

The Schwarzschild radius $R$ is given by

\begin{eqnarray}
R&=&\tau \sinh r\nonumber\\
&=&{3\over K}\sinh {Kl\over 3}
\label{2.3}
\end{eqnarray}
and its derivative reads

\begin{equation}
R'   =   {dR\over dl}=\cosh {Kl\over 3}.
\label{2.4}
\end{equation}

The primary objects we deal with are the optical scalars, the expansion
$\theta $  of the outgoing null rays and the convergence $\theta '$ of
the ingoing light rays. These are given by

\begin{eqnarray}
R\theta   & =& 2R' +{2\over 3}KR\nonumber\\
&=&2\cosh {Kl\over 3} + 2\sinh {Kl\over 3}\nonumber\\
&=& 2e^{Kl\over 3}
\label{2.5}
\end{eqnarray}
and similarly

\begin{eqnarray}
R\theta '= 2 e^{-Kl\over 3},
\label{2.6}
\end{eqnarray}
therefore the product of $R\theta$ and $R\theta '$ remains constant,
$R\theta R\theta '=4$. We also have

\begin{eqnarray}
R\theta = {4RK\over 3}+2 e^{-Kl\over 3}.
\label{2.7}
\end{eqnarray}

Thus at the origin we have

\begin{eqnarray}
R\theta '=R\theta =2
\label{2.8}
\end{eqnarray}
and at infinity one of the scalars is divergent while the other vanishes

\begin{eqnarray}
&&R\theta   \rightarrow {4\over 3}KR, \nonumber\\
&&R\theta '   \rightarrow  0.
 \label{2.9}
\end{eqnarray}

An alternative form of the metric (\ref{2.1}) is that in terms of the
Schwarzschild
radius

 \begin{equation}
ds^2 = {-\tau^2\over \tau^2+R^2}  d\tau^2-{2R\tau \over \tau^2+R^2} dR
d\tau + {\tau^2\over  \tau^2+R^2}  dR^2 +   R^2  d\Omega^2.
\label{2.10}
\end{equation}
\section{GENERAL STRUCTURE OF THE SPHERICALLY SYMMETRIC CONSTRAINTS}
\label{sec3}

The two divergences of null rays are given by

\begin{equation}
 \omega_+= R\theta=Rp-RK^r_r+RK,
\label{omegap}
\end{equation}
\begin{equation}
 \omega_-= R\theta'=Rp+RK^r_r-RK,
\label{omegam}
\end{equation}
where
\begin{equation}
p=\frac{2} {R}\frac{dR}{dl},
\label{p}
\end{equation}
is the mean curvature of a surface of constant $R$ in the slice where $R$ is
the
Schwarzschild radius and $l$ is the proper distance.
The constraints now can be written as

\begin{equation}
\partial_l(\omega_+)= -8\pi R(\rho-j) -\frac{1}{
4R}(2\omega_+^2-\omega_+\omega_-
-4-4\omega_+RK),
\label{domegap}
\end{equation}

\begin{equation}
\partial_l(\omega_-)= -8\pi R(\rho+j) -{1\over 4R}(2\omega_-^2-\omega_+\omega_-
-4+4\omega_-RK),
\label{domegam}
\end{equation}

\begin{equation}
\partial_lR=R'=\frac{1}{4}(\omega_++\omega_-).
\label{dR}
\end{equation}

We assume we are given $\rho$ (the energy density), $j=\vec{ j}\cdot \hat n$
(the
current density), where $\hat n$ is the outgoing radial normal and $R K$
 as functions of $l$ and then solve the triplet of O.D.E's (\ref{domegap}),
(\ref{domegam}) and  (\ref{dR}) for $(R,\omega_+,\omega_-)$.
The only conditions we assume are regularity at the origin
$(R=0,\omega_+=\omega_-=2) $, asymptotic flatness  and that the sources satisfy
the dominant
energy condition, $\rho\geq |j|$.

Combining Eq. (\ref{domegap}) and (\ref{domegam}) we can write
\begin{equation}
\partial_l(\omega_-\omega_+)= -8 \pi R\big (\rho(\omega_++\omega_-)+
j(\omega_+-\omega_-)\big )-{1\over 4R}(\omega_-\omega_+-4)(\omega_++\omega_-)
\label{domegapm}
\end{equation}
and by regularity and  asymptotic flatness we have that
$\lim_{R\to 0}\omega_-\omega_+=4$ also
$\lim_{R\to \infty}\omega_-\omega_+=4$.

Suppose that $ \omega_-\omega_+>4$, if both are positive we have that the right
hand side of Eq. (\ref{domegapm}) is strictly negative and if both are negative
the right
hand side is positive. Thus we have

\begin{equation}
\omega_-\omega_+\leq 4.
\label{estomega}
\end{equation}

 \section{CONSTRAINTS ON ASYMPTOTICALLY FLAT CMC HYPERSURFACES. }
\label{sec4}

 When we consider asymptotically flat CMC hypersurfaces, it is useful to use
variables that are finite at the origin and infinity. From the Minkowski
analysis, it is clear that we need as boundary conditions that
$\omega_+\rightarrow 2e^{Kl/3}$  and $\omega_-\rightarrow 2e^{-Kl/3}$.
Thus the natural variables to use are $A= \omega_+e^{-Kl/3}$ and
$B=\omega_-e^{Kl/3}$. Using these the equations (\ref{domegap}) and
(\ref{domegam}) become

\begin{equation}
\partial_lA=-8\pi Re^{-Kl/3}(\rho -j)
-{e^{Kl/3}\over 4R}
[2A^2-{8\over 3}KRe^{-Kl/3}A-ABe^{-2Kl/3} -4e^{-2Kl/3}],
\label{4.1}
\end{equation}

\begin{equation}
\partial_lB=-8\pi Re^{Kl/3}(\rho +j)-{e^{-Kl/3}\over 4R}
[2B^2+{8\over 3}KRe^{Kl/3}B-ABe^{2Kl/3} -4e^{2Kl/3}].
\label{4.2}
\end{equation}

We know, from the previous Section (inequality (\ref{estomega})) that
$AB=\omega_+ \omega_-$ is bounded above
by 4 .
Let us write the expression which does not depend on the sources in Eq.
(\ref{4.1}) as

\begin{equation}
-{e^{Kl/3}\over 2R}[A^2-{4\over 3}KRe^{-Kl/3}A  -4e^{-2Kl/3}]
-{e^{-Kl/3}\over 4R}[4-AB].
\label{4.3}
\end{equation}

Consider

\begin{eqnarray}
&&\alpha = 2\sqrt{{K^2R^2\over 9}+1} +{2\over 3}KR , \nonumber\\
&&\beta = 2\sqrt{{K^2R^2\over 9}+1} -{2\over 3}KR , \nonumber\\
\label{4.4}
\end{eqnarray}
these are essentially the roots of the  quadratic equation in $A$ in
(\ref{4.3}). If $A$ lies outside the range

\begin{equation}
[-\beta e^{-Kl/3}, \alpha e^{-Kl/3}],
\label{4.5}
\end{equation}
then every term on the right hand side of (\ref{4.1}) is negative and
therefore $\partial_lA<0$.

It is clear that $\alpha > {4\over 3}KR$ and also $\alpha \ge 2$, and these
are the limiting values of $A$ at infinity and at the origin respectively.
At any maximum of $A$ we have that   $\partial_lA=0$ which implies that
$A\le  \alpha e^{-Kl/3}$ at that point. We can show that this is a global
upper bound, or equivalently

\begin{equation}
\omega_+\le  2\sqrt{{K^2R^2\over 9}+1} +{2\over 3}KR.
\label{4.6}
\end{equation}

Consider the function

\begin{equation}
f(l)= 2\sqrt{{K^2R^2\over 9}+1} +{2\over 3}KR -\omega_+,
\label{4.7}
\end{equation}
$f(l)$ is zero at the origin and by asymptotic flatness it is positive at
infinity. We can show that
it is always positive. Let us assume, to the contrary, that $f(l)$ is negative
somewhere. This means that there must exist a negative minimum, i. e., a
point where
\begin{itemize}
\item[(i)]
 $ \omega_+>  2\sqrt{{K^2R^2\over 9}+1} +{2\over 3}KR$ and
\item[(ii)]
 $f'(l)=0$.
\end{itemize}

However, we can show that if (i) holds then  $f'(l)>0$. Using Eq.
(\ref{domegap})
and (\ref{dR}), we can calculate

\begin{eqnarray}
f'(l)&=& {1\over 4R}\bigg [{KR\over 3}
{\alpha (\omega_++\omega_-)\over \sqrt{{K^2R^2\over 9}+1}}\nonumber\\
& &+2\omega_+^2-\omega_+\omega_--4\omega_+KR-4\bigg ]
+8\pi R(\rho -j).
\label{4.8}
\end{eqnarray}

The coefficient of $\omega_-$ is

\begin{equation}
{1\over 4R}\bigg [-\omega_+ +{KR\alpha \over 3\sqrt{{K^2R^2\over 9}+1}}\bigg ].
\label{4.9}
\end{equation}

This is obviously negative since $\omega_+\ge \alpha $.
Therefore we minimize $f'(l)$ by choosing the maximum value of $\omega_-$.
There is a condition, Eq. (\ref{estomega}), that constrains the
product of both optical scalars, $\omega_+\omega_- \leq 4$. Therefore the
maximum of
$\omega_-$ is ${4\over \omega_+}$.

Now consider the function

\begin{eqnarray}
\tilde f(l)&=& {1\over 4R}\bigg [{KR\over 3}
{\alpha (\omega_++{4\over \omega_+})\over \sqrt{{K^2R^2\over 9}+1}}\nonumber\\
& &+2\omega_+^2 -8-4\omega_+KR \bigg ].
\label{4.10}
\end{eqnarray}

The derivative of this function with respect to $\omega_+$ is positive
(assuming $\omega_+ \ge \alpha $) and therefore its minimum is achieved
at the minimum of $\omega_+$, i. e., at $\omega_+=\alpha $.

Hence

\begin{eqnarray}
 f'(l)&\geq& {1\over 4R}\bigg [{KR\over 3}
{\alpha^2+4  \over \sqrt{{K^2R^2\over 9}+1}}
 +2 \alpha^2 -8 -4 \alpha KR \bigg ]\nonumber\\
&+&8\pi R(\rho -j).
\label{4.11}
\end{eqnarray}

It is easy to see that this expression is positive. Hence we get a
contradiction. We could show, in a similar vein, the existence of the
upper bound on $B$ as well of the lower bounds.
Therefore we have the following global bounds on the optical
scalars

\begin{equation}
 -\beta \leq \omega_+ \leq \alpha ,
\label{4.12}
\end{equation}

\begin{equation}
 -\alpha \leq \omega_- \leq \beta .
\label{4.13}
\end{equation}

Those bounds are valid for both signs, positive and negative, of the trace of
the extrinsic curvature $K$.
\section{SUFFICIENT CONDITION FOR TRAPPED SURFACES IN CMC HYPERSURFACES}
\label{sec5}
We can use the formulae (\ref{domegap}) and (\ref{dR}) in Section \ref{sec2} to
derive
\begin{equation}
\partial_l(\omega_+R)= -8\pi R^2(\rho-j)+1 +\frac{1}{
4}(2\omega_+\omega_--\omega_+^2
+4\omega_+RK)
\label{dRomegap}
\end{equation}
and we  have the bounds on $\omega_+$ and $\omega_-$ from the previous section.
 It is easy to show that the maximum value of $(2\omega_+\omega_--
\omega_+^2+4\omega_+RK)$ occurs when i) $\omega_+=\alpha$ and $\omega_-=\beta$
if $K > )$, and ii) $\omega_+=-\beta$ and $\omega_-=-\alpha$.
Hence we get in both cases

\begin{equation}
2\omega_+\omega_--\omega_+^2+4\omega_+RK \leq
4+{16\over 9}K^2R^2+{16\over 3}|K|R\sqrt{{K^2R^2\over 9}+1}.
\label{est1}
\end{equation}

We can complete the square in the square root to finally get
\begin{equation}
{1\over 4}(2\omega_+\omega_--\omega_+^2+4\omega_+RK)\leq 1+{8\over 9}K^2R^2+
{4\over 3}|K|R,
\label{est2}
\end{equation}
thus  we get
\begin{equation}
\partial_l(\omega_+R)\leq -8\pi R^2(\rho-j)+2+{8\over 9}K^2R^2+{4\over 3}|K|R.
\label{estdRomegap}
\end{equation}

If we integrate this equation out to some surface $S$ we get
\begin{equation}
\omega_+R|_S\leq -2(M-P)+2L+{2K^2\over 9\pi}V+{4\over 3}|K|\int_0^lRdl,
\label{estRomegap}
\end{equation}
where $M=\int 4\pi R^2\rho dl$ is the total amount of matter inside $S$,
$P=\int 4\pi R^2j dl$ is the total outward radial momentum of the matter,
$L$ is the proper radius and $V$ is the volume inside $S$ (notice that
$4\pi R^2\rho dl=dV$ is  the proper volume element). Therefore we have that if

$$
(M-P)(S)\geq L+{K^2\over 9\pi}V+{2\over 3}|K|\int Rdl,
$$
we must have that $\omega_+R|_S$ is negative and so the surface  at $S$ is a
future
trapped surface. We can estimate $\int R dl$ as
follows:
\begin{equation}
\int R dl\leq \bigg [\int R^2dl\bigg ]^{1\over 2}\bigg [\int dl\bigg ]^{1\over
2}=
\bigg ({VL\over 4\pi}\bigg )^{1\over 2}.
\label{suff}
\end{equation}

Therefore a sufficient condition  for the appearance of a future trapped
surface on a slice with constant trace of the extrinsic curvature  is
that
\begin{equation}
(M-P)(S)\geq L+{K^2\over 9\pi}V+{|K|\over 3 }\bigg ({VL\over \pi}
\bigg )^{1\over 2}.
\label{suffcond}
\end{equation}

\section{A NECESSARY CONDITION FOR A TRAPPED SURFACE IN A CMC HYPERSURFACE}

Let us return to the equality  (\ref{dRomegap}) we derived in Section
\ref{sec5}
and again integrate it out to some surface S

\begin{equation}
\omega_+R|_S= -2(M-P)+L(S)+{1\over 4}\int_0^L(2\omega_+\omega_--
\omega_+^2+4\omega_+RK) dl,
\label{Romegap}
\end{equation}
but now we wish to minimize the integral rather than maximize it. We assume
that
no future trapped surface exists within $S$ , i.e. $\omega_+\geq 0$. We also
 assume that no past trapped surface exists in $S$. Not only that, but that
 $\omega_-$ is strongly  bounded away from zero, i.e. $\omega_-\geq C>0$; that
means that all radially ingoing null rays are converging.

In other words we want to minimize the function
$f(\omega_+,\omega_-)=2\omega_+\omega_--
\omega_+^2+4\omega_+RK$ in the region given by $0\leq\omega_+\leq\alpha$ and
 $C\leq\omega_-\leq\beta$.  A simple calculation gives
$f_{\text{min}}=\min(f(\alpha,C),0)$.

Because $\alpha $ is a function of $R$, we need to study  the function

\begin{eqnarray*}
\tilde f(R)&=&f(\alpha(R),C)\nonumber\\
&=&2\alpha C-\alpha^2+4\alpha R K
\end{eqnarray*}
in order to find $f_{\text{min}}$.

We will consider separately two cases, with the positive and negative trace
of the extrinsic curvature.

i) Let $K>0$. By inspection  we obtain that
this function is an increasing function in the variable $R$, therefore
$$
\min(\tilde f)= f(\alpha(0),C)=4C-4.
$$

Clearly when $C\geq 1$ $f_{\text{min}}=0$ otherwise $f_{\text{min}}\geq 4C-4 $.
Inserting this into
 Eq. \ref{Romegap} we obtain two estimates

\begin{equation}
\omega_+R|_S\geq \left \{ \begin{array}{ll}
			   -2(M-P)(S)+L(S) &\mbox{for $ C\geq 1$}\\
			    & \\
			   -2(M-P)(S)+CL(S) &\mbox{for $ C\leq 1$,}
			   \end{array}
		 \right .
\label{Romegalow}
\end{equation}
that is, since $\theta(S)=0$,

$$
M(S)-P(S)\geq \left \{ \begin{array}{ll}
			{L\over 2} &\mbox{for $ C\geq 1$}\\
					      & \\
			{CL\over 2}&\mbox {for $  C\leq 1$}
			\end{array}
	       \right .
$$
is the necessary condition for the existence of a trapped surface.

The above result obviously applies to maximal slices. In connection with that,
two of us have to admit that Theorem 2 in \cite{EMNOM} should be stated
as above; the actual statement of \cite{EMNOM} that the necessary condition
for future trapped surfaces is $M(S)-P(S)\geq {CL\over 2}$ can be wrong.

ii) Let $K<0$.  In this case one easily estimates $\tilde f(R)$ from below
by
$$
 \left \{ \begin{array}{ll}
			4(C-1)+ {4KR(4 + C)\over 3 }  &\mbox{for $ C\leq 1$}\\
					      & \\
			 {4KR(4 + C)\over 3} &\mbox {for $  C\geq 1$}
			\end{array}
	       \right .
$$
That leads to  a necessary condition for $S$ to be trapped
$$
M(S)-P(S)+{(4 + C)|K|\over 6}\int_0^{L(S)}dlR(l)\geq \left \{
\begin{array}{ll}
			{L\over 2} &\mbox{for $ C\geq 1$}\\
					      & \\
			{CL\over 2}&\mbox {for $  C\leq 1$}
			\end{array}
	       \right .
$$
Using   relations $dl={2\over Rp}dR$ and   $pR={1\over 2}
(\omega_++\omega_-)\ge {C\over 2}$  one obtains
$$ \int_0^{L(S)}dlR(l)\le {S\over 2\pi C}$$
and  the necessary condition
$$
M(S)-P(S)+ {(4 + C)|K|S\over 12\pi C}\geq \left \{ \begin{array}{ll}
			{L\over 2} &\mbox{for $ C\geq 1$}\\
					      & \\
			{CL\over 2}&\mbox {for $  C\leq 1$}
			\end{array}
	       \right .
$$
A similar necessary condition, under a somewhat stronger condition,
has been obtained by Zannias (\cite{zannias}). Thus negative values of the
trace of the extrinsic curvature can help to form trapped surfaces.

Let us recall that yet another necessary result has been derived in
\cite{emnompen}, where the following equation has been proven
\begin{eqnarray}
&&{R^3\over 8}\theta (S)\theta '(S)+m-{S\over 16\pi }^{1/2}= \nonumber\\
&&\pi \int_r^{\infty }\sqrt{a}R^3\Bigl[ \rho_0(\theta +\theta ')
+j (\theta -\theta ')\Bigr]. \nonumber\\
\label{penrose}
\end{eqnarray}
(\ref{penrose}) has been derived on maximal slices, but it  holds true on any
slicing, assuming a quick enough falloff of matter fields. Under the
dominant energy condition one concludes that an outermost trapped surface $S$
(future or past) must have a radial radius $R$ not greater than $2m$.
This conclusion is slicing-independent.

\section{REISSNER - NORDSTR\"OM GEOMETRY  IN CMC FOLIATIONS}
\label{sec7}

In this Section we will present an explicit line element for
 electrovacuum      in constant mean curvature foliations.
The most general spherically symmetric line element can be put
\begin{equation}
ds^2=-N^2dt^2+adr^2+R^2d\Omega^2.
\label{7.1}
\end{equation}

We assume that the trace of the extrinsic curvature
\begin{equation}
K={\partial_t(\sqrt{a}R^2)\over 2N\sqrt{a}R^2}
\label{7.2}
\end{equation}
is constant on a particular slice and, morever, is time independent.
The three nonzero components of the extrinsic curvature are
\begin{equation}
K_r^r={\partial_t(\sqrt{a} )\over 2N\sqrt{a} },~~~K_{\phi }^{\phi }=
K_{\theta }^{\theta }={\partial_tR\over NR }={1\over 2}(K-K_r^r).
\label{7.3}
\end{equation}

The spherically symmetric Einstein equations consist of constraint equations
(\ref{constraint}), the evolution equation
\begin{equation}
\partial_t(K_r^r-trK)=-{p^3R^2\over 2N}{\partial_r\over \sqrt{a} }
({ N\over pR })^2+{3N\over 2}(K_r^r)^2 +8\pi (T_r^r+\rho )+{N\over 2}K^2
-2NKK_r^r
\label{7.4}
\end{equation}
and the lapse equation
\begin{equation}
\Delta^{(3)}N=
 N[{3\over 2}(K_r^r)^2 +{1\over 2}K^2-KK_r^r +4\pi (T_i^i+\rho )].
\label{7.5}
\end{equation}

In electrovacuum  we have ${q^2\over 8\pi R^4}=\rho = T_i^i=-T^r_r$, where $q$
is
the electric charge.
The mean curvature $p$ of nested two spheres and the extrinsic curvatures
are easily found from the constraints (\ref{constraint}) and they read
\begin{eqnarray}
&&pR=2\sqrt{1+{C\over R}+{q^2\over R^2} +({KR\over 3}+ {C_1\over 2R^2})^2},
\nonumber\\
&&K_r^r={K\over 3}+{C_1\over R^3}. \nonumber\\
\label{7.6}
\end{eqnarray}

The lapse equation becomes now
\begin{equation}
\Delta^{(3)}N=
 N[{3C^2_1\over 2R^6}+{q^2\over R^4} +{K^2\over 3}]
\label{7.7}
\end{equation}
and one easily finds out that it is solved by
\begin{equation}
 N= \gamma {pR \over 2},
\label{7.8}
\end{equation}
where $\gamma $ is given by
\begin{equation}
\gamma(r,t) =1+ C_2 \int_{R(r)}^{\infty }d\tilde R {1\over (\tilde R)^2(p
\tilde R)^3}.
\label{7.9}
\end{equation}

Inserting the whole information into (\ref{7.4}) and using the relation
${\partial_r\over \sqrt{a}}={Rp\over 2}\partial_R$ one obtains
that the constant $ C_2$ depends on the rate of change of the radial
- radial component of the extrinsic curvature,

\begin{equation}
C_2 =4\partial_tC_1.
 \label{7.10}
\end{equation}

The change of the coordinate variable $r$ into the areal radius $R$
transforms the line element (\ref{7.1}) into

\begin{eqnarray}
ds^2&=&dt^2[-N^2+{\gamma^2\over 4}\Bigl(  {C_1\over R^2}-{2KR\over 3} \Bigr)
^2]
+2\gamma {{C_1\over R^3}-{2K\over 3}\over p}dtdR+ {4\over (pR)^2}dR^2+
R^2d\Omega^2,\nonumber\\
&=&-{\gamma^2\omega_+\omega_-\over 4}dt^2+2\gamma{\omega_--\omega_+\over
  \omega_-+\omega_+}dRdt+{16\over (\omega_-+\omega_+)^2 }dR^2+R^2d\Omega^2,
\label{7.11}
\end{eqnarray}

with $N$ and $p$ defined above. Let us point out that the parameter $C$
that appears in the expression for $p$ may be identified with $-2m_B$,
where $m_B$ is the Bondi mass.

\section{ ``CMC SURFACES AVOID SINGULARITIES''}

In the article \cite{EMNOM} an argument was advanced as to how foliations with
bounded trace of the extrinsic curvature might avoid singularities. In this
section we wish to produce a different (and sharper) argument to the same end.
This argument works for essentially any slicing, but we present it
here specifically for CMC slices. Let us consider a model of a collapsing
system
where the support of the matter becomes ever smaller as the collapse continues
so that eventually the star is confined to a region much smaller than that
enclosed by the apparent horizon. If the star were to be compressed inside a
boundary which satisfies $R \ll m$, where $m$ is the conserved ADM mass of the
star, before any singularity appears then regular CMC foliations will be
excluded from this part of the spacetime.

{}From the inequalities (\ref{4.12}) and (\ref{4.13}) we can show that on any
regular CMC slice
\begin{eqnarray}
\omega_+\omega_- &\geq& -4({2K^2R^2 \over 9} + 1 + {2|K|R \over 3}\sqrt{1 +
K^2R^2 \over 9})\nonumber\\
&\geq& -4\bigl({2|K|R \over 3} + 1\bigr)^2. \label{8.1}
\end{eqnarray}
However, both
the Schwarzschild radius
 $R$, and the product $\omega_+\omega_-$ are four-scalars and we have
\begin{equation}
\omega_+\omega_-= 4(1 - {2m_{H} \over R}). \label{8.2}
\end{equation}
where $m_{H}$ is the so-called Hawking mass, which equals the constant ADM mass
outside  the support of the matter. Inequality (\ref{8.1}) and equality
(\ref{8.2})  can be combined to give the following inequality
\begin{equation}
m_{H} \le {2K^2R^3 \over 9} + {2|K|R \over 3} + R.\label{8.3}
\end{equation}
This means that for a fixed positive $m_{H}$ we have a lower bound for $R$. Let
us assume that we are considering a spherical collapse and viewing it using
a CMC foliation. Let us also assume that during this collapse a two-surface
appears which violates this inequality (\ref{8.3}) before the CMC slices become
singular. This means that the CMC foliation cannot progress past this point,
the
lapse collapses. Since the lapse equation is elliptical, not only does the
lapse
go to zero at this point, it becomes small on the whole interior and the CMC
slicing freezes.

This means that if we wish to find a solution where CMC slices run right up to
the singularity we cannot allow a large accumulation of matter near the center
before the singularity appears. A possible way for this to happen is that if
the collapse were such that in addition to the infall of matter, one also had
an explosion that pushed significant amounts of the star outwards, away from
the horizon.

This bound is valid for solutions which have the spatial
topology $R^1 \times S^2$ as in the extended Schwarzschild solution as well as
topology $R^3$.
 Maximal slicing can be viewed as a special case
of CMC slicing and it was observed many years ago (see \cite{bryce}) that the
regular maximal slicing of the Schwarzschild solution saturates at $R = 3m/2$,
in
agreement with the bound stated above.

\acknowledgments
This work was initiated during the ESI Summer School in Mathematical
Relativity,
Vienna, 1994. It has been partially supported by the Forbairt grant
SC/94/225. and the KBN project 2 PO3B 090 08.

\end{document}